\documentclass[5p,twocolumn]{elsarticle}

\usepackage{hyperref}
\usepackage{graphicx}
\graphicspath{ {images/} }

\journal{arXiv}

\biboptions{authoryear}

\begin{document}

\begin{frontmatter}

\title{Benchmarking confound regression strategies for the control of motion artifact in studies of functional connectivity}

\author[neuropsych]{Rastko Ciric}
\author[neuropsych]{Daniel H. Wolf}
\author[cornell]{Jonathan D. Power}
\author[neuropsych]{David R. Roalf }
\author[neuropsych]{Graham Baum}
\author[neuropsych]{Kosha Ruparel}
\author[stats]{Russell T. Shinohara}
\author[radiology]{Mark A. Elliott}
\author[dusseldorf,julich]{Simon B. Eickhoff}
\author[radiology]{Christos Davatzikos}
\author[neuropsych]{Ruben C. Gur}
\author[neuropsych]{Raquel E. Gur}
\author[be,ee]{Danielle S. Bassett}
\author[neuropsych]{Theodore D. Satterthwaite \corref{mycorrespondingauthor}}
\ead{sattertt@mail.med.upenn.edu}

\cortext[mycorrespondingauthor]{Corresponding author}
\address[neuropsych]{Department of Psychiatry, Perelman School of Medicine, University of Pennsylvania, Philadelphia PA, USA}
\address[cornell]{Sackler Institute for Developmental Psychobiology, Weill Medical College of Cornell University, New York, NY, USA}
\address[stats]{Department of Biostatistics and Epidemiology, Perelman School of Medicine, University of Pennsylvania, Philadelphia, PA, USA}
\address[radiology]{Department of Radiology, Perelman School of Medicine, University of Pennsylvania, Philadelphia PA, USA}
\address[dusseldorf]{Department of Clinical Neuroscience and Medical Psychology, Heinrich-Heine University D{\"u}sseldorf}
\address[julich]{Institute of Neuroscience and Medicine (INM-1), Research Center J{\"u}lich}
\address[be]{Department of Bioengineering, University of Pennsylvania, Philadelphia PA, USA}
\address[ee]{Department of Electrical and Systems Engineering, University of Pennsylvania, Philadelphia PA, USA}

\begin{abstract}
Since initial reports regarding the impact of motion artifact on measures of functional connectivity, there has been a proliferation of confound regression methods to limit its impact. However, recent techniques have not been systematically evaluated using consistent outcome measures. Here, we provide a systematic evaluation of 12 commonly used confound regression methods in 193 young adults. Specifically, we compare methods according to three benchmarks, including the residual relationship between motion and connectivity, distance-dependent effects of motion on connectivity, and additional degrees of freedom lost in confound regression. Our results delineate two clear trade-offs among methods. First, methods that include global signal regression minimize the relationship between connectivity and motion, but unmask distance-dependent artifact. In contrast, censoring methods mitigate both motion artifact and distance-dependence, but use additional degrees of freedom. Taken together, these results emphasize the heterogeneous efficacy of proposed methods, and suggest that different confound regression strategies may be appropriate in the context of specific scientific goals.
\end{abstract}


\end{frontmatter}


\section*{Introduction}

Resting-state (intrinsic) functional connectivity (rsfc-MRI) has evolved to become one of the most common brain imaging phenotypes \citep{Craddock2013,Fox2007,Power2014b,Smith2013,VanDijk2010}, and has been critical for understanding fundamental properties of brain organization \citep{Damoiseaux2006,Fox2005,Power2011,Yeo2011}, brain development over the lifespan \citep{DiMartino2014,Dosenbach2011,Fair2008}, and abnormalities associated with diverse clinical conditions \citep{Baker2014,Buckner2008,Fair2010}. rsfc-MRI has numerous advantages, including ease of acquisition and suitability for a wide and expanding array of analysis techniques. However, despite knowledge that in-scanner motion can influence measures of activation from task-related fMRI \citep{Friston1996}, the impact of in-scanner motion on measures of functional connectivity was not explored for 16 years after its initial discovery. However, since the near-simultaneous publication of three independent reports in early 2012 \citep{VanDijk2012,Power2012,Satterthwaite2012}, it has been increasingly recognized that motion can have a large impact on rsfc-MRI measurements, and can systematically bias inference. This bias is particularly problematic in developmental or clinical populations where motion is correlated with the independent variable of interest (age, diagnosis) \citep{Satterthwaite2012,Fair2012}, and has resulted in the re-evaluation of numerous published findings.

In response to this challenge, there has been a recent proliferation of methods aimed at mitigating the impact of motion on functional connectivity \citep{Yan2013a,Power2015}. These methods can be broadly grouped into several categories. First, high-parameter confound regression strategies use expansions of realignment parameters or tissue-compartment signals, often including derivative and quadratic regressors \citep{Friston1996,Satterthwaite2013,Yan2013a}. Second, principal component analysis (PCA) based methods (CompCor; \citet{Behzadi2007,Muschelli2014}) find the primary directions of variation within high-noise areas defined by anatomy (e.g., aCompCor) or temporal variance (tCompCor). Third, whole-brain independent component analysis (ICA; \citet{Beckmann2005}) of single-subject time series has increasingly been used for de-noising, with noise components selected either by a trained classifier (ICA-FIX; \citet{Griffanti2014,Salimi-Khorshidi2014}) or using \textit{a priori} heuristics (ICA-AROMA; \citet{Pruim2015a,Pruim2015b}). Fourth, temporal censoring techniques identify and remove (or de-weight) specific volumes contaminated by motion artifact, often followed by interpolation. These techniques include scrubbing \citep{Power2012,Power2014a,Power2015}, spike regression \citep{Satterthwaite2013}, and de-spiking \citep{Jo2013,Patel2014}. Censoring techniques have been reported to attenuate motion artifact, but at the cost of a shorter time series and variably reduced degrees of freedom. Fifth, one recent report emphasized the relative merits of spatially-tailored confound regression using local white matter signals (wmLocal; \citet{Jo2013}). Finally, the inclusion of global signal regression (GSR) \citep{Macey2004} in confound regression models remains a source of controversy \citep{Fox2009,Murphy2009,Chai2012,Saad2012,Yan2013b}. While several studies have suggested its utility in de-noising \citep{Fox2009,Power2015,Satterthwaite2013,Yan2013a}, other studies have emphasized the risk of removing a valuable signal \citep{Yang2014,Hahamy2014}, potentially biasing group differences \citep{Gotts2013,Saad2012}, or exacerbating distance-dependent motion artifact. Distance-dependent artifact \citep{Power2012,Satterthwaite2012} manifests as increased connectivity in short-range connections, and reduced connectivity in long-range connections, which has the potential to impact measures of network topology \citep{Yan2013c}. 

This recent proliferation of de-noising techniques has prompted excitement but also sowed confusion. Unsurprisingly, new de-noising pipelines have often tended to emphasize outcome measures that suggest their relative superiority. As a result, investigators often anecdotally report substantial uncertainty regarding which pipeline should be used. Such uncertainty has been exacerbated by the lack of common outcome measures used across studies, which has hampered direct comparison among pipelines. While one review paper has summarized recent developments in this rapidly-evolving sub-field \citep{Power2015}, systematic evaluation of de-noising pipelines according to a range of benchmarks remains lacking. 

Accordingly, in this report we compare a dozen of the most commonly used confound regression strategies in a large ($N = 193$) dataset of young adults. Pipelines evaluated include standard techniques, high-parameter confound regression, PCA-based techniques such as aCompCor and tCompcor, ICA-based approaches such as ICA-AROMA, spatially-tailored local white matter regression, and three different censoring techniques (spike regression, de-spiking, and scrubbing); GSR is included in many pipelines as well. Critically, we compare these pipelines according to three intuitive benchmarks, including the residual relationship between functional connectivity and subject motion, the degree of distance-dependent artifact, and the loss of temporal degrees of freedom. As described below, results underscore the relative strengths and weaknesses among these methods, and outline clear trade-offs among commonly used confound regression approaches. 

\section*{Materials and methods}

\subsection*{Participants and data acquisition}

The task-free BOLD data used in this study ($N = 193$) were selected from the Philadelphia Neurodevelopmental Cohort (PNC) \citep{Satterthwaite2014,Satterthwaite2016} on the basis of age, health, and data quality. In order to minimize the potential impact of developmental effects, subjects selected for inclusion were aged at least 18 years at the date of scan. All participants were free of medical problems that could impact brain function \citep{Merikangas2010}, lacked gross structural brain abnormalities \citep{Gur2013}, and were not taking psychotropic medication.  Furthermore, subjects were excluded from the study if they failed to satisfy any of three criteria for functional image quality: if subject movement (relative root mean square displacement) averaged over all frames exceeded 0.2mm; if over 20 individual frames featured movement in excess of 0.25mm \citep{Satterthwaite2012}; or if brain coverage during acquisition was incomplete. As a result of these criteria, participants with gross in-scanner motion were excluded, allowing us to evaluate the utility of confound regression strategies for the mitigation of artifact due to micro-movements.

Structural and functional subject data were acquired on a 3T Siemens Tim Trio scanner with a 32-channel head coil (Erlangen, Germany), as previously described \citep{Satterthwaite2014,Satterthwaite2016}. High-resolution structural images were acquired in order to facilitate alignment of individual subject images into a common space. Structural images were acquired using a magnetization-prepared, rapid-acquisition gradient-echo (MPRAGE) T1-weighted sequence ($T_{R} = 1810$ms; $T_{E} = 3.51$ms; FoV $= 180\times240$mm; resolution 1mm isotropic). Approximately 6min of task-free functional data were acquired for each subject using a blood oxygen level-dependent (BOLD-weighted) sequence ($T_{R} = 3000$ms; $T_{E} = 32$ms; FoV $= 192\times192$mm; resolution 3mm isotropic; 124 spatial volumes). Prior to scanning, in order to acclimate subjects to the MRI environment and to help subjects learn to remain still during the actual scanning session, a mock scanning session was conducted using a decommissioned MRI scanner and head coil. Mock scanning was accompanied by acoustic recordings of the noise produced by gradient coils for each scanning pulse sequence. During these sessions, feedback regarding head movement was provided using the MoTrack motion tracking system (Psychology Software Tools, Inc, Sharpsburg, PA). Motion feedback was only given during the mock scanning session. In order to further minimize motion, prior to data acquisition subjects' heads were stabilized in the head coil using one foam pad over each ear and a third over the top of the head. During the resting-state scan, a fixation cross was displayed as images were acquired. Subjects were instructed to stay awake, keep their eyes open, fixate on the displayed crosshair, and remain still.

\subsection*{Structural image processing}

A study-specific template was generated from a sample of 120 PNC subjects balanced across sex, race, and age bins using the buildTemplateParallel procedure in ANTs \citep{Avants2011a}. Study-specific tissue priors were created using a multi-atlas segmentation procedure \citep{Wang2014}. Next, each subject's high-resolution structural image was processed using the ANTs Cortical Thickness Pipeline \citep{Tustison2014}. Following bias field correction \citep{Tustison2010}, each structural image was diffeomorphically registered to the study-specific PNC template using the top-performing SyN deformation \citep{Klein2009}. Study-specific tissue priors were used to guide brain extraction and segmentation of the subject's structural image \citep{Avants2011b}.

\subsection*{BOLD time series processing}

Task-free functional images were processed using the XCP Engine (Ciric et al., In Preparation), which was configured to support the 12 pipelines evaluated in this study (see \textbf{Figure 1}). Each pipeline was based on de-noising strategies previously described in the neuroimaging literature. A number of preprocessing procedures were included across all de-noising pipelines. Common elements of preprocessing included (1) correction for distortions induced by magnetic field inhomogeneities, (2) removal of the 4 initial volumes of each acquisition, (3) realignment of all volumes to a selected reference volume using \textsc{mcflirt} \citep{Jenkinson2002}, (4) demeaning and removal of any linear or quadratic trends, (5) co-registration of functional data to the high-resolution structural image using boundary-based registration \citep{Greve2009}, and (6) temporal filtering using a first-order Butterworth filter with a passband between 0.01 and 0.08 Hz. These preliminary processing stages were then followed by the confound regression procedures described below. In order to prevent frequency-dependent mismatch during confound regression \citep{Hallquist2013}, all regressors were band-pass filtered to retain the same frequency range as the data. As in our prior work \citep{Satterthwaite2012,Satterthwaite2013}, the primary summary metric of subject motion used was the mean relative root-mean-square displacement calculated during time series realignment using \textsc{mcflirt}.

\subsection*{Overview of confound regression strategies}

The primary objective of the current study was to evaluate the performance of common de-noising strategies. We selected 12 de-noising models, labelled \textbf{1}--\textbf{12} below, for evaluation (Figure \ref{fig:pipelines}). Models \textbf{1}--\textbf{5} used nuisance parameters derived from 6 movement estimates and 3 physiological time series, as well as their temporal derivatives and quadratic expansions.

\begin{itemize}
\item \textbf{Model 1.} (2P) Used only the 2 physiological time series: mean signal in WM and mean signal in CSF, and functioned as a base model for comparison to other more complex confound regression models.
\item \textbf{Model 2.} (6P) Used only the 6 motion estimates derived from \textsc{mcflirt} realignment as explanatory variables.
\item \textbf{Model 3.} (9P) Combined the 6 motion estimates and 2 physiological time series with global signal regression. This model has been widely applied to functional connectivity studies \citep{Fox2005,Fox2009}.
\item \textbf{Model 4.} (24P) Expansion of model \textbf{2} that includes 6 motion parameters, 6 temporal derivatives, 6 quadratic terms, and 6 quadratic expansions of the derivatives of motion estimates for a total 24 regressors \citep{Friston1996}.
\item \textbf{Model 5.} (36P) Similar expansion of model \textbf{3}: 9 regressors, their derivatives, quadratic terms, and squares of derivatives \citep{Satterthwaite2013}.
\end{itemize}

Models \textbf{6}--\textbf{8} further expanded upon this maximal 36P strategy by incorporating censoring approaches.

\begin{itemize}
\item \textbf{Model 6.} (36P+despike) Included 36 regressors as well as despiking \citep{Cox1996}.
\item \textbf{Model 7.} (36P+spkreg) Included 36 regressors as well as spike regression, as in \citet{Satterthwaite2013}.
\item \textbf{Model 8.} (36P+scrub) Included 36 regressors as well as motion scrubbing, as in \citet{Power2014a}.
\end{itemize}

Models \textbf{9} and \textbf{10} adapted variants of the PCA-based \textit{CompCor} approach.

\begin{itemize}
\item \textbf{Model 9.} (aCompCor) Used 5 principal components each from the WM and CSF, in addition to motion estimates and their temporal derivatives \citep{Muschelli2014}.
\item \textbf{Model 10.} (tCompCor) Used 6 principal components from high-variance voxels \citep{Behzadi2007}.
\end{itemize}

The final two models evaluated used a regressor derived from local WM signals, and subject-specific ICA de-noising, respectively.  

\begin{itemize}
\item \textbf{Model 11.} (wmLocal) Used a voxelwise, localised WM regressor in addition to motion estimates and their temporal derivatives and despiking \citep{Jo2013}.
\item \textbf{Model 12.} (ICA-AROMA) Used a recently developed ICA-based procedure for removal of motion-related variance from BOLD data, together with mean WM and CSF regressors \citep{Pruim2015b,Pruim2015a}.
\end{itemize}

We explicitly limited our scope to models that did not require training a classifier, and did not evaluate confound regression strategies that require extensive parameter optimization \citep{Salimi-Khorshidi2014,Griffanti2014,Patel2014}. Furthermore, in order to constrain the parameter space, we did not examine unpublished combinations of de-noising approaches.

\subsubsection*{Confound regression using realignment parameters} \label{tsreg}

Time series of six realignment parameters (three translational and three rotational) for each subject were returned by \textsc{mcflirt} as part of time series realignment (motion correction). Additionally, the temporal derivative, quadratic terms, and quadratic of the temporal derivative of each of the realignment parameters were calculated, yielding 24 realignment regressors in total. The original six realignment parameters were included in confound regression models \textbf{2} and \textbf{3}.  Models \textbf{9} and \textbf{11} included 12 realignment regressors -- the 6 realignment parameters and their temporal derivatives -- while the full set of 24 expanded realignment regressors were included as part of confound regression models \textbf{4}--\textbf{8}.

\subsubsection*{Global signal regression}

The mean global signal was computed by averaging across all voxelwise time series located within a subject-specific mask covering the entire brain. The global signal was included in model \textbf{3}, while the expanded models \textbf{5}--\textbf{8} included 4 global signal regressors: the global signal, its derivative, its square, and the derivative of its square.

\subsubsection*{Tissue class regressors}

Mean white matter (WM) and cerebrospinal fluid (CSF) signals were computed by averaging within masks derived from the segmentation of each subject's structural image; these masks were eroded using AFNI's 3dmask\_tool \citep{Cox1996} to prevent inclusion of gray matter signal via partial-volume effects. The WM mask was eroded at the 2-voxel level, while the CSF mask was eroded at the 1-voxel level. More liberal erosion often resulted in empty masks in our data. Temporal derivatives, quadratic terms, and squares of the derivative were computed as above. Two tissue class regressors (WM and CSF)  were included in models \textbf{3} and \textbf{12}, whereas their expansions (8 regressors) were included in models \textbf{4}--\textbf{8}.

\subsubsection*{Local white matter regression}

Model \textbf{11} used a local WM regressor \citep{Jo2013}. This was computed in AFNI using 3dLocalstat \citep{Cox1996}. Unlike the regressors described above, which were voxel-invariant, the value of the local WM regressor was computed separately at each voxel. For each voxel, a sphere of radius 45mm was first centered on that voxel; this sphere defined that voxel's local neighborhood. Next, this spherical neighborhood was intersected with an eroded WM mask to produce a local WM mask, which included only the fraction of the WM that was also in the voxel's neighborhood. The mean signal within this new local WM mask was then used to model the local WM signal at the voxel \citep{Jo2013}. This process was repeated at every voxel in order to generate the local WM regressor. This local WM regressor was included in model \textbf{11} along with realignment parameters and their derivatives (12 total); this model also included voxelwise de-spiking.

\subsubsection*{CompCor}

Principal component analysis (PCA) can be used to model noise in BOLD time series \citep{Behzadi2007,Muschelli2014}. Broadly, the use of PCA-derived regressors to model noise is called component-based correction (CompCor). Numerous variants of CompCor have been developed; here, our focus will be on anatomical CompCor (aCompCor, model \textbf{9}) and temporal CompCor (tCompCor, model \textbf{10}). In aCompCor, a PCA is performed within an anatomically defined tissue class of interest. We extracted 5 components for WM and CSF each, yielding 10 compcor components \citep{Muschelli2014}. As part of model \textbf{9}, as in \citet{Muschelli2014},  these 10 aCompCor components were combined with 12 re-alignment parameters (raw and temporal derivative). In tCompCor, the temporal variance of the BOLD signal is first computed at each voxel. Subsequently, a mask is generated from high-variance voxels, and principal components are extracted from the time series at these voxels. In confound regression model \textbf{10}, tCompcor was implemented using ANTs, with 6 tCompCor components used as confound regressors for each participant. 

\subsubsection*{ICA-AROMA}

ICA-AROMA (automatic removal of motion artifact) is a recently-introduced, widely-used method for de-noising using single-subject ICA \citep{Pruim2015b,Pruim2015a}; we evaluated ICA-AROMA in confound regression model \textbf{12}. In contrast to other ICA based methods (e.g., ICA-FIX: \citet{Salimi-Khorshidi2014}), it does not require dataset-specific training data. The input to ICA-AROMA is a voxelwise time series that has been smoothed at 6mm FWHM using a Gaussian kernel. After decomposing this time series using FSL's \textsc{melodic} (with model order estimated using the Laplace approximation) \citep{Beckmann2005}, ICA-AROMA uses four features to determine whether each component corresponds to signal or noise. The first 2 features are spatial characteristics of the signal source: (1) the fraction of the source that falls within a CSF compartment and (2) the fraction of the source that falls along the edge or periphery of the brain. The remaining features are derived from the time series of the source: (3) its maximal robust correlation with time series derived from realignment parameters and (4) its high-frequency spectral content. ICA-AROMA includes two de-noising steps. The first de-noising step occurs immediately after classification. All component time series (signal and noise) are included as predictors in the linear model, and the residual BOLD time series is obtained via partial regression of only the noise time series. A second confound regression step occurs after temporal filtering, wherein mean signals from WM and CSF were regressed from the data.

\subsubsection*{Temporal censoring: de-spiking, spike regression, and scrubbing}

In addition to regression of nuisance time series, a number of `temporal censoring' approaches were used to identify motion-contaminated volumes in the BOLD time series and reduce their impact on further analysis. These approaches included despiking, spike regression, and scrubbing. Despiking is a procedure that identifies outliers in the intensity of each voxel's detrended BOLD time series and then interpolates over these outliers. Despiking was implemented in AFNI using the 3dDespike utility \citep{Cox1996} as part of confound regression model \textbf{6}.

Unlike despiking, which identifies outliers on a voxelwise basis, spike regression and scrubbing censor complete volumes based on metrics of subject movement defined \textit{a priori}. For spike regression, as in \citet{Satterthwaite2013}, volumes were flagged for spike regression if their volume-to-volume RMS displacement exceeded 0.25mm. Next, as part of confound regression model \textbf{7}, $k$ `spike' regressors were included as predictor variables in the de-noising model, where $k$ equalled the number of volumes flagged \citep{Satterthwaite2013}. For each flagged time point, a unit impulse function that had a value of 1 at that time point and 0 elsewhere was included as a spike regressor.

For scrubbing, the framewise displacement (FD) \citep{Power2012} was computed at each time point as the sum of the absolute values of the derivatives of translational and rotational motion estimates. If framewise displacement (FD) at any point in time exceeded 0.2mm, then that time point was flagged for scrubbing. It should be noted that the conversion of FD to RMS displacement is approximately 2:1, and thus the published criterion for scrubbing has a lower threshold for flagging high-motion volumes than does spike regression. Scrubbing of BOLD data was performed iteratively \citep{Power2014a} as part of confound regression model \textbf{8}. At any stage where a linear model was applied to the data (for instance, during detrending procedures), high-motion epochs were temporally masked out of the model so as not to influence fit. During temporal filtering, a frequency transform was used to generate surrogate data with the same phase and spectral characteristics as the unflagged data. This surrogate data was used to interpolate over flagged epochs prior to application of the filter. During confound regression, flagged timepoints were excised from the time series so as not to contribute to the model fit. For scrubbing (but not spike regression) if fewer than five contiguous volumes had unscrubbed data, these volumes were scrubbed and interpolated as well.

\subsection*{Overview of outcome measures}

We evaluated each de-noising pipeline according to three benchmarks. Residual \textit{QC-FC correlations} and \textit{distance-dependence} provided a metric of each pipeline's efficacy, while \textit{loss of temporal DOF} provided an estimate of each pipeline's efficiency.

\subsubsection*{QC-FC correlations}

In order to estimate the residual relationship between subject movement and connectivity after de-noising, we computed \textit{QC-FC} correlations (quality control / functional connectivity) \citep{Power2015,Satterthwaite2012,Satterthwaite2013,Power2012}. While other metrics have been used in prior reports, including FD-DVARS correlations, we favor \textit{QC-FC} as the most useful metric of interest as it directly quantifies the relationship between motion and the primary outcome of interest (rather than two quality metrics, as in FD-DVARS). For an extended discussion of the rationale for this measure, see \citet{Power2015}.

We evaluated \textit{QC-FC} relationships within two commonly-used whole-brain networks, the first consisting of spherical nodes distributed across the brain \citep{Power2011} and the second comprising an areal parcellation of the cerebral cortex \citep{Gordon2016}. For each network, the mean time series for each node was calculated from the denoised residual data, and the pairwise Pearson correlation coefficient between node time series was used as the network edge weight \citep{Biswal1995}. For each edge, we then computed the correlation between the weight of that edge and the mean relative RMS motion. To eliminate the potential influence of demographic factors, \textit{QC-FC} relationships were calculated as partial correlations that accounted for participant age and sex. We thus obtained, for each de-noising pipeline, a distribution of \textit{QC-FC} correlations. This distribution was used to obtain two measures of the pipeline's ability to mitigate motion artifact, including: 1) the number of edges significantly related to motion, which was computed after using the false discovery rate (FDR) to account for multiple comparisons; and 2) the median absolute value of all \textit{QC-FC} correlations. All graphs were generated using ggplot2 in R version 3.2.3; brain renderings were prepared in BrainNet Viewer \citep{Xia2013}.

\subsubsection*{Distance-dependent effects of motion}

Early work on motion artifact demonstrated that in-scanner motion can bias connectivity estimates between two nodes in a manner that is related to the distance between those nodes \citep{Satterthwaite2012,Power2012}. Under certain processing conditions, subject movement enhances short-distance connections while reducing long-distance connections. To determine the residual distance-dependence of subject movement, we first used the center of mass of each node to obtain a distance matrix $D$ where entry $D_{ij}$ indicates the Euclidean distance between the centers of mass of nodes $i$ and $j$. We then obtained the correlation between the distance separating each pair of nodes and the \textit{QC-FC} correlation of the edge connecting those nodes; this correlation served as an estimate of the distance-dependence of motion artifact.

\subsubsection*{Additional degrees of freedom lost in confound regression}

Including additional regressors in a confound model reduces the impact of motion on future analyses, but it is not without cost. Confound regressors and censoring both reduce the temporal degrees of freedom (DOF) in data. This loss in temporal DOF may diminish the ability of functional data acquisitions to sample a subject's connectome or may introduce bias if the loss is variable across subjects. Thus, de-noising strategies ideally limit the loss of temporal DOF, for instance by including fewer, more efficacious regressors. In the present study, we also assessed the number of temporal DOF lost in each confound regression approach.

As in previous work \citep{Pruim2015b}, we assumed that each time series regressed out and each volume excised from the data constituted a single temporal DOF. Consequently, the loss of temporal DOF was estimated as the sum of the number of regressors in each confound model and the number of volumes flagged for excision under that model. It should be emphasized that the values thus obtained are imperfect estimates. First, because functional MR time series typically exhibit temporal autocorrelation, the actual loss in DOF will be less than the estimated loss in DOF. Accordingly, censoring adjacent volumes does not remove the same number of DOF as does censoring volumes separated in time. Furthermore, a temporal bandpass filter was uniformly applied to all data prior to confound regression; this filtering procedure would itself have removed additional temporal DOF and elevated the autocorrelation of the data. Because this filter was uniform across all de-noising strategies, it was not considered when estimating the loss of additional temporal DOF in each model.

\section*{Results}

\subsection*{Heterogeneity in confound regression performance}

Confound regression strategies typically remove some, but not all, of the artifactual variance that head motion introduces into the BOLD signal. The motion-related artifact that survives de-noising can be quantified to provide a metric of pipeline performance.  Here, our primary benchmark of confound regression efficacy was the residual relationship between brain connectivity and subject motion, or the \textit{QC-FC correlation}. We measured \textit{QC-FC} correlations using two metrics: the percentage of network connections where a significant relationship with motion was present (\textbf{Figure \ref{fig:nse}}), and the absolute median correlation (AMC) between connection strength and head movement across all connections (\textbf{Figure \ref{fig:amc}}).

No preprocessing strategy was completely effective in abolishing the relationship between head movement and connectivity. However, different approaches exhibited widely varying degrees of efficacy.  The top four confound regression strategies included 36 parameters, comprising an expansion of GSR, tissue-specific regressors (WM, CSF), and realignment parameters. Beyond this base 36-parameter model, all censoring techniques provided provided some additional benefit, reducing the number of edges that were significantly related to motion to less than 1\%. Convergent results were present across both \textit{QC-FC} measures (\% edges, AMC) and networks (Power, Gordon) that were evaluated.

In contrast, many pipelines performed relatively poorly, leaving a majority of network edges with a residual relationship with motion. Specifically, 82\% of edges were impacted by motion when the least effective method was used (6 realignment parameters). The commonly used 24-parameter expansion of realignment parameters originally suggested by \citet{Friston1996} did not provide much of an improvement (79\% edges).  Similarly, the  local WM regressor model (69\% edges) and tCompCor (48\% edges) also resulted in substantial residual \textit{QC-FC} correlations. In fact, these methods performed worse than a basic 2-parameter model composed of mean WM and CSF signals (33\% edges). Finally, several methods demonstrated intermediate performance, with 1-20\% of edges impacted by motion.  This middle group included methods as disparate as aCompCor (5\% edges), ICA-AROMA (12\% edges), and the classic 9-parameter confound regression model which included GSR (4\% edges).

\subsection*{Variability in distance-dependent motion artifact after confound regression}

Our second benchmark quantified the distance-dependent motion artifact that was present in data processed by each pipeline (\textbf{Figure \ref{fig:distdep}}). We observed that distance-dependence was present even under conditions where artifact magnitude was attenuated. For example, though the 36-parameter model was among the most effective in attenuating \textit{QC-FC} relationships, its application revealed strongly distance-dependent artifact. Examination of graphs that plot \textit{QC-FC} by Euclidean distance (see \textbf{Figure \ref{fig:distdep}C}) revealed that this is due to effective mitigation of motion artifact for long-range but not short-range connections.

Distance-dependence was most prominent in models that included GSR, but did not include censoring (e.g., 9-parameter and 36-parameter models). However, despite the lack of global signal in the aCompCor or tCompcor models, data returned from both of these component-based approaches revealed substantial distance-dependent artifact. Notably, inclusion of censoring consistently reduced distance-dependence, although scrubbing was somewhat more effective than spike regression or voxelwise despiking.

The top performing method according to this benchmark was ICA-AROMA, which completely abolished any distance-dependence of residual motion artifact. In other words, the motion artifact that was still present in the data after ICA-AROMA impacted all connections in a manner that was not dependent on the spatial separation between nodes. There was similar lack of distance-dependence in the wmLocal model, although as noted above this model did not provide effective de-noising according to \textit{QC-FC} benchmarks. Despite the presence of the global signal, the 36-parameter model that included scrubbing also performed quite well.

\subsection*{Effective preprocessing strategies use many additional degrees of freedom}

Perhaps unsurprisingly, the preprocessing strategies that consistently reduced both \textit{QC-FC} correlations and distance-dependence were also among the costliest in terms of loss of temporal degrees of freedom (\textbf{Figure \ref{fig:dof}}). By definition, the 36-parameter models included a high fixed number of regressors. Furthermore, models that additionally included censoring resulted in a substantial additional loss of data that varied across subjects.  ICA-AROMA also had a variable loss of DOF, but of a lower magnitude than censoring or high-parameter confound regression. 

\section*{Discussion}

In response to rapid evolution of confound regression strategies available for the mitigation of motion artifact, in this report we evaluated 12 commonly-used pipelines. Results indicate that there is substantial heterogeneity in the performance of these confound regression techniques across all measures evaluated. The context, implications, and limitations of these findings are discussed below.

\subsection*{Confound regression techniques have substantial performance variability}

We evaluated confound regression strategies according to three intuitive benchmarks that were selected to capture different domains of effectiveness. These included \textit{QC-FC} associations, distance-dependence of motion artifact, and additional degrees of freedom lost in confound regression. Across each benchmark, there was a striking heterogeneity in pipeline performance.

Notably, five of the top six confound regression approaches included GSR. This effect was consistent in both networks we evaluated. The effectiveness of GSR is most likely due to the nature of motion artifact itself: in-scanner head motion tends to induce widespread reductions in signal intensity across the entire brain parenchyma (see \citet{Satterthwaite2013}, Figure 4). As discussed in detail elsewhere (Power et al., Under Revision) this effect is highly reproducible across datasets, and is effectively captured by time series regression of the global signal.

Beyond GSR, a second strategy that clearly minimizes \textit{QC-FC} relationships is temporal censoring. We evaluated three censoring variants, including scrubbing, spike regression, and de-spiking.  Compared to spike regression and de-spiking, scrubbing appears to be more effective in removing distance-dependent artifact in this dataset. This is most likely due to the explicit tension between data quality and data quantity: because of the lower threshold for scrubbing than spike regression (due to differences in FD vs. RMS measures of motion; see Figure 9C in \citet{Yan2013a}), more low-quality data was excised during scrubbing.  Furthermore, scrubbing includes a criterion to not leave isolated epochs ($<5$ volumes) of un-scrubbed data.  Consequently, this leads to clear differences in the additional degrees of freedom lost by each method. In contrast to spike regression and scrubbing, which eliminate high motion volumes completely, time series de-spiking identifies and interpolates large changes in signal intensity on a voxelwise basis \citep{Cox1996}. This allows for spatial adaptivity (see \citet{Patel2014}) but also renders quantification of data loss and comparisons with volume-based censoring techniques more difficult.

Critically, while both GSR and censoring appeared effective in minimizing \textit{QC-FC} relationships, they exhibited opposite effects on distance-dependence. While censoring techniques appear to consistently reduce the presence of distance-dependence, GSR is associated with increased distance-dependence. However, it should be emphasized that the distance-dependence associated with GSR is not the result of worsening associations with motion in certain connections. Rather, the distance-dependence seen with GSR stems from differential de-noising efficacy, whereby motion artifact is more effectively minimized for long-distance connections than for short-range connections. Certain models such as the local WM regression approach \citep{Jo2013} thus have minimal distance-dependence, but this is a consequence of lack of efficacy across all distances. In contrast, ICA-AROMA \citep{Pruim2015a,Pruim2015b} reduced motion to a moderate degree over all connection distances, resulting in almost no distance-dependence. However, while clearly an improvement over some other methods, data processed using ICA-AROMA was noisier than other methods which included GSR or censoring, and resulting networks contained a substantial number of edges impacted by motion. 
Somewhat to our surprise, benchmark results for aCompCor \citep{Behzadi2007,Muschelli2014} were most similar to models that included GSR.  Alone among models where GSR was not included, aCompCor was both relatively effective in the mitigation of residual motion (5\% of edges impacted) and also exhibited substantial distance-dependence (e.g., $r=-0.26$). This suggests that while aCompCor does not explicitly include GSR, the practical results of its application are in fact quite similar.

\subsection*{Trade-offs of confound regression approaches: implications for investigators}

The current results emphasize two clear trade-offs in the choice of confound regression strategy. First, pipelines that include global signal regression tend to be more effective at minimizing \textit{QC-FC} relationships, but at the cost of some increase in distance-dependence. As noted above, for minimizing \textit{QC-FC} relationships, nearly all of the top strategies (except aCompCor) included GSR. Conversely, the two techniques that had the most substantial distance-dependence (the 9-regressor and 36-regressor methods) both included GSR. Second, censoring techniques provide a clear benefit in reducing \textit{QC-FC} relationships and additionally tend to attenuate distance-dependence. However, by definition, removing contaminated volumes results in less data and loss of degrees of freedom.

These two trade-offs suggest that a single confound regression strategy is unlikely to be optimal for every study. For example, in studies of network organization, the presence of both anti-correlations and distance-dependent artifact resulting from GSR may result in an undesirable impact on nodal degree distribution \citep{Yan2013c}. In contrast, for studies of group or individual differences, minimizing \textit{QC-FC} relationships is likely to be of paramount importance so as to limit the possibility that inference is driven by artifactual signals. This concern is particularly relevant for studies of development or clinical sub-groups where motion is systematically related to the subject-level variable of interest (e.g., age, disease status).  Co-varying for motion at the group level is unlikely to be a panacea for such studies when inadequate subject-level time series de-noising is employed, as prior work \citep{Power2014a} has suggested that motion effects at the group level may potentially both be nonlinear and vary across sub-samples in a manner that is difficult to predict. However, aggressive volume censoring may be problematic in datasets with relatively brief acquisitions. In datasets where long time series are acquired, such as multi-band acquisitions \citep{Feinberg2010} and intensive acquisitions of single subjects \citep{Laumann2015}, loss of temporal degrees of freedom is less likely to be a major concern. The 36-parameter models without volume censoring offer uniformity, as does randomly or systematically censoring additional volumes until all subjects retain approximately the same degrees of freedom.

\subsection*{Limitations}

Several limitations of the current approach should be noted. One of the principal challenges in evaluating the performance of de-noising approaches is the lack of a noise-free ground truth. Our primary benchmark of confound regression performance assumes that mitigation of the relationship between QC (participant motion) and FC (i.e., the imaging measurement) is desirable. To the degree that in-scanner motion itself represents a biologically informative phenotype, this approach will mistake signal for noise. Indeed, prior data suggests that this may sometimes be the case. For example, \citet{Zeng2014} found specific changes in connectivity for participants who had generally high levels of motion, even on scans where motion was low. However, without multiple scans which allow such careful dissociation, most studies are incapable of disambiguating the large confounding effects of motion on connectivity. Second, in place of \textit{QC-FC} relationships, one could focus on alternative benchmarks such as test-retest reliability \citep{Zuo2014}. Reliability is certainly of interest, but to the degree that motion tends to be highly correlated within individuals across scanning sessions, there is a substantial potential for the presence of consistent motion artifact across sessions to artificially inflate reliability, and diminish the biological relevance of observed results. A third and related concern is that certain de-noising methods could conceivably both minimize \textit{QC-FC} relationships and even enhance reliability by aggressively removing both signal and noise, but in the process diminish sensitivity to meaningful individual differences. Indeed, one prior study demonstrated the association between canonical resting state networks and randomly generated confound parameters \citep{Bright2015}. This concern is somewhat mitigated by other studies in independent datasets, which suggest that higher-order confound regressors improve the confound regression model fit \citep{Yan2013a,Satterthwaite2013}, while random regressors do not (see Figure 8 in \citet{Satterthwaite2013}). Nonetheless, the tension between the goals of noise reduction and sensitivity to individual differences remains an outstanding issue for the field. Fourth, while our evaluation included many of the most commonly used techniques, other approaches which require substantial training or parameter selection (i.e., ICA-FIX \citep{Salimi-Khorshidi2014,Griffanti2014}, wavelet de-spiking \citep{Patel2014}) may be valuable and merit further consideration. Fifth and finally, it should be noted that while improvements in image acquisition (including multi-echo techniques) may not salvage existing motion-contaminated data, it is likely that they will change the methodological landscape of connectivity research moving forward \citep{Kundu2012,Kundu2013,Bright2013}.

\subsection*{Conclusions}

Taken together, the present results underline the performance heterogeneity of recently-introduced, commonly-used confound regression methods. In selecting among these methods, investigators should be aware of the relative strengths and weaknesses of each approach, and understand how processing strategy may impact inference. Clearly, the relative merit of each approach will vary by research question and study design. Perhaps most importantly, as has been emphasized in nearly every other study of motion artifact, the choice of confound regression strategy is often dwarfed in importance by the need to transparently report and evaluate the impact of motion in each dataset. At a minimum, this includes reporting the relationship between motion artifact and not only subject phenotypes (e.g., group, age, symptom or cognitive score) but also the functional connectivity phenotypes being considered. In the context of such data, the distinction between observed results and the impact of motion artifact can be understood. Such transparency bolsters confidence in reported findings, but also will likely tend to emphasize the remaining challenges for de-noising going forward. Especially when considered in the context of the the rapid evolution of available techniques since 2012, there is no doubt that innovations in post-processing confound regression strategies will continue.

\section*{Acknowledgements}

Thanks to the acquisition and recruitment team, including Karthik Prabhakaran and Jeff Valdez. Thanks to Chad Jackson for data management and systems support. Thanks to Monica Calkins for phenotyping expertise. Supported by grants from the National Institute of Mental Health: R01MH107703 (TDS), R01MH107235 (RCG), and R01NS089630 (CD). The PNC was funded through NIMH RC2 grants MH089983, and MH089924 (REG). Additional support was provided by R21MH106799 (DSB \& TDS), R01MH101111 (DHW), K01MH102609 (DRR), P50MH096891 (REG), R01NS085211 (RTS), and the Dowshen Program for Neuroscience. DSB acknowledges support from the John D. and Catherine T. MacArthur Foundation, the Alfred P. Sloan Foundation, the Army Research Laboratory and the Army Research Office through contract numbers W911NF-10-2-0022 and W911NF-14-1-0679, the National Institute of Mental Health (R01-DC-009209-11), the National Institute of Child Health and Human Development (R01HD086888-01), the Office of Naval Research, and the National Science Foundation (BCS-1441502 and PHY-1554488). Support for developing statistical analyses (RTS \& TDS) was provided by a seed grant by the Center for Biomedical Computing and Image Analysis (CBICA) at Penn.  Data deposition: The data reported in this paper have been deposited in database of Genotypes and Phenotypes (dbGaP), www.ncbi.nlm.nih.gov/gap (accession no. phs000607.v1.p1). 

\section*{References}

\bibliography{mybibfile}

\clearpage{}
\newgeometry{margin=1.75cm}
\begin{figure*}
    \centering
    \includegraphics[width=\textwidth]{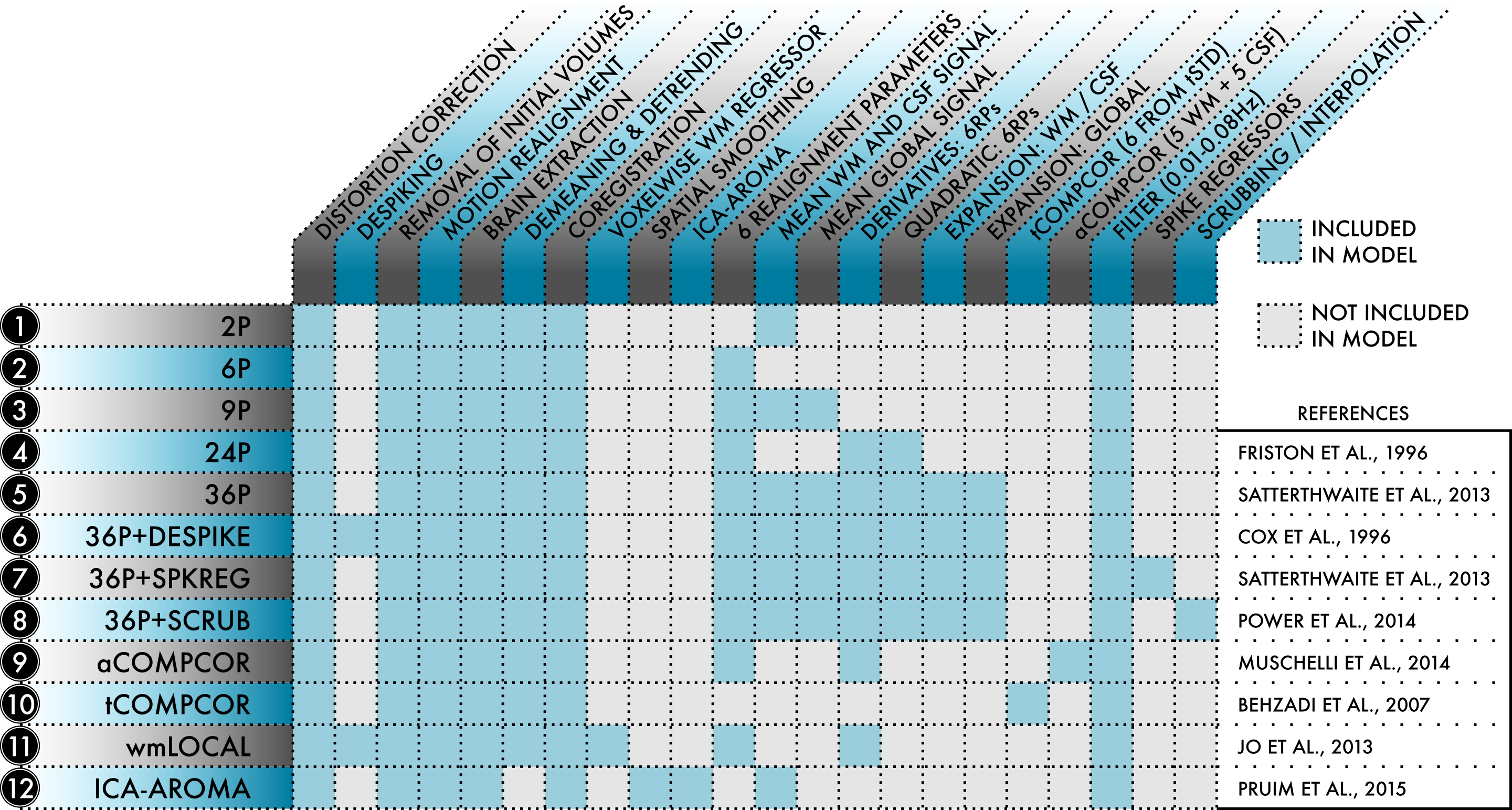}
    \caption{\textbf{Schematic of the 12 de-noising models evaluated in the present study.} For each of the 12 models indexed at left, the table details what processing procedures and confound regressors were included in the model. De-noising models were selected from the functional connectivity literature and represented a range of strategies.}
    \label{fig:pipelines}
\end{figure*}

\clearpage{}
\newgeometry{margin=1.75cm}
\begin{figure*}
    \centering
    \includegraphics[width=\textwidth]{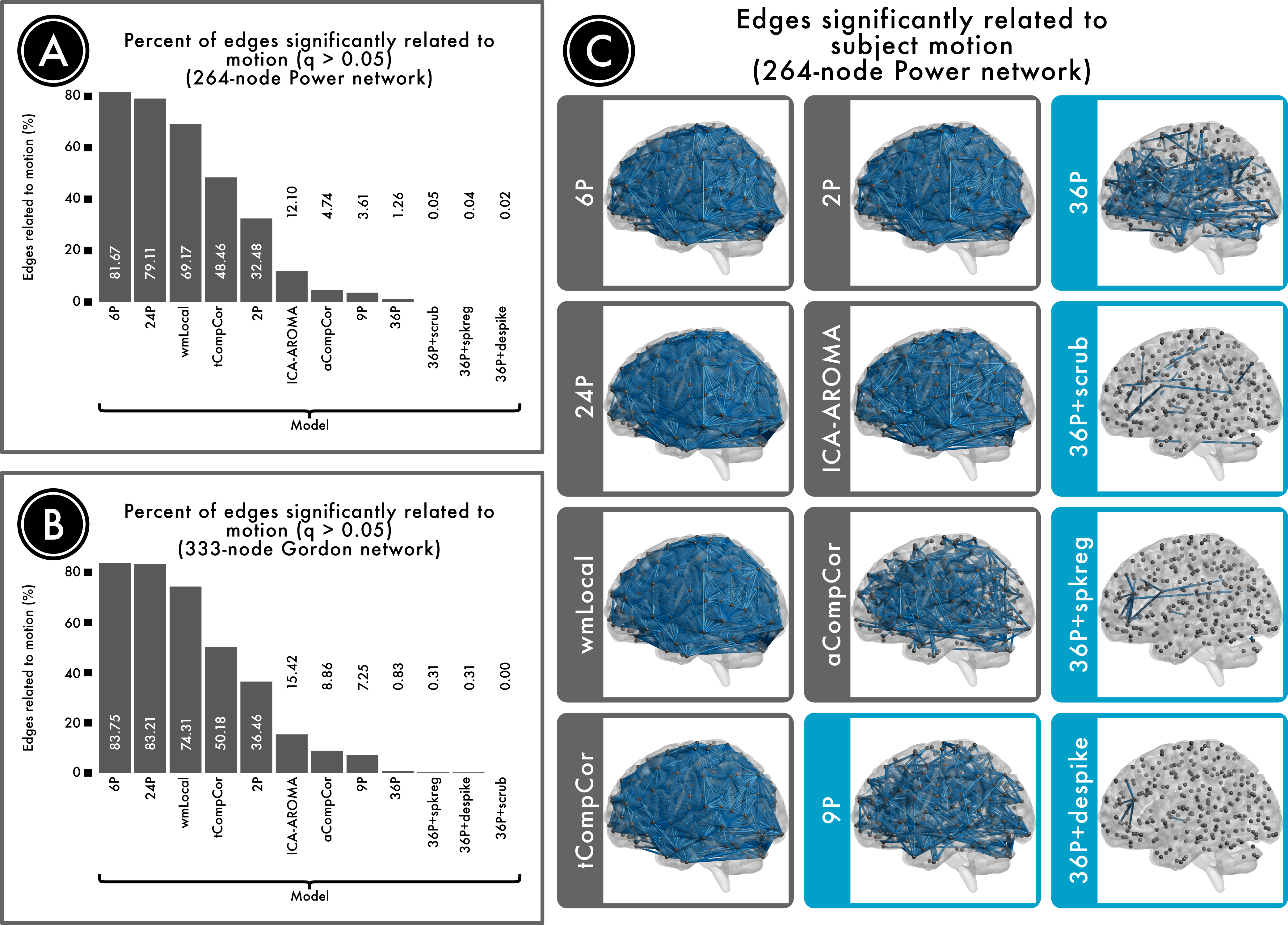}
    \caption{\textbf{Number of edges significantly related to motion after de-noising.} Successful de-noising strategies reduced the relationship between connectivity and motion. The number of edges (network connections) for which this relationship persists provides evidence of a pipeline's efficacy. \textbf{\textit{A,}} The percentage of edges significantly related to motion in a 264-node network defined by \citet{Power2011}. Fewer significant edges is indicative of better performance. \textbf{\textit{B,}} The percentage of edges significantly related to motion in a second, 333-node network defined by \citet{Gordon2016}. \textbf{\textit{C,}} Renderings of significant edges for each de-noising strategy, ranked according to efficacy. Strategies that include regression of the mean global signal are framed in blue and consistently ranked as the best performers.}
    \label{fig:nse}
\end{figure*}

\clearpage{}
\newgeometry{margin=1.75cm}
\begin{figure*}
    \centering
    \includegraphics[width=\textwidth]{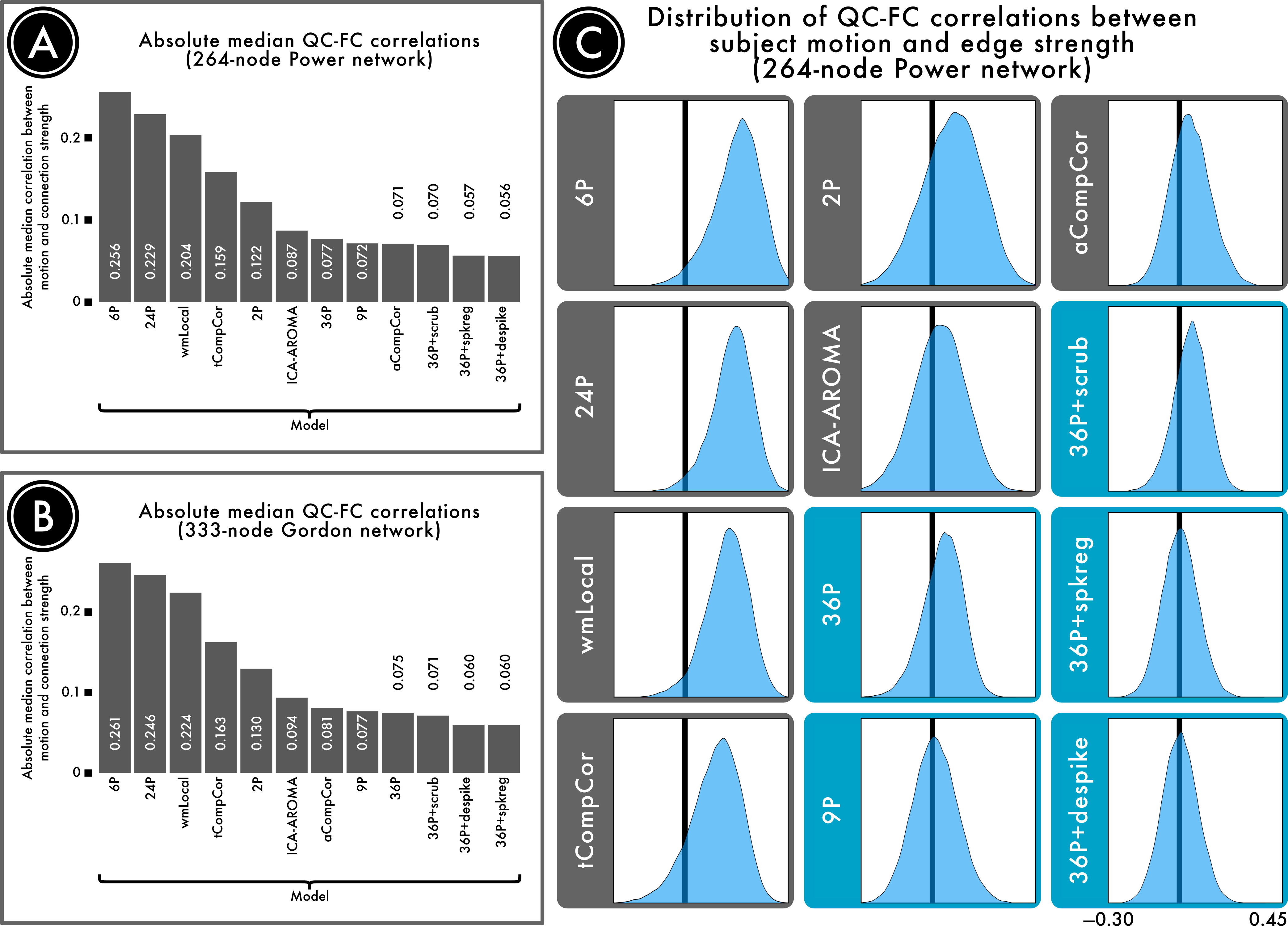}
    \caption{\textbf{Residual QC-FC correlations after de-noising.} The absolute median \textit{QC-FC} correlation is another measure of the relationship between connectivity and motion. \textbf{\textit{A,}} The absolute median correlation between functional connectivity and motion in a 264-node network defined by \citet{Power2011}. A lower absolute median correlation indicates better performance. \textbf{\textit{B,}} The absolute median correlation between functional connectivity and motion in a second, 333-node network defined by \citet{Gordon2016}. \textbf{\textit{C,}} Distributions of all edgewise \textit{QC-FC} correlations after each de-noising strategy, ranked according to efficacy. Results largely recapitulated those reported in Figure \ref{fig:nse}, with GSR-based approaches (blue frame) collectively exhibiting the best performance. Whereas approaches that included more regressors generally yielded a narrower distribution, those approaches that included GSR tended to shift the distribution's center toward 0.}
    \label{fig:amc}
\end{figure*}

\clearpage{}
\newgeometry{margin=1.75cm}
\begin{figure*}
    \centering
    \includegraphics[width=\textwidth]{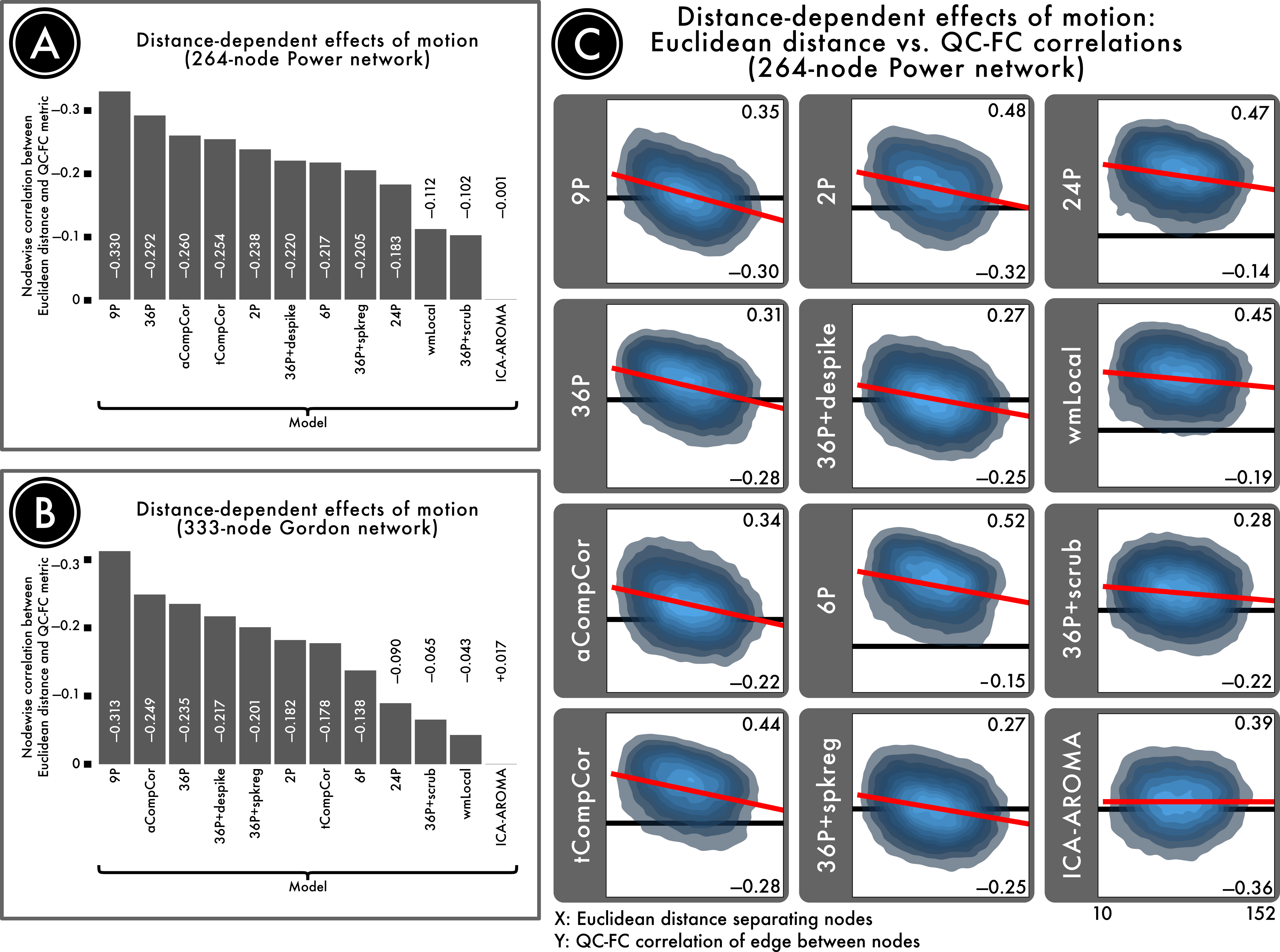}
    \caption{\textbf{Distance-dependence of motion artifact after de-noising.} The magnitude of motion artifact varies with the Euclidean distance separating a pair of nodes, with closer nodes generally exhibiting greater impact of motion on connectivity. \textbf{\textit{A,}} The residual distance-dependence of motion artifact in a 264-node network defined by \citet{Power2011} following confound regression. \textbf{\textit{B,}} The residual distance-dependence of motion artifact in a second, 333-node network defined by \citet{Gordon2016}. \textbf{\textit{C,}} Density plots indicating the relationship between the Euclidean distance separating each pair of nodes (x-axis) and the \textit{QC-FC} correlation of the edge connecting those nodes (y-axis). The overall trend lines for each de-noising strategy, from which distance-dependence is computed, are indicated in red. The best performing models either excised high-motion volumes (36-parameter + scrubbing) or  used more localized regressors (ICA-AROMA and wmLocal). In general, approaches that made use of GSR without censoring resulted in substantial distance-dependence. This effect was driven by differential efficacy of de-noising, with effective de-noising for long range connections but not short-range connections.}
    \label{fig:distdep}
\end{figure*}

\clearpage{}
\newgeometry{margin=1.75cm}
\begin{figure*}
    \centering
    \includegraphics[width=0.5\textwidth]{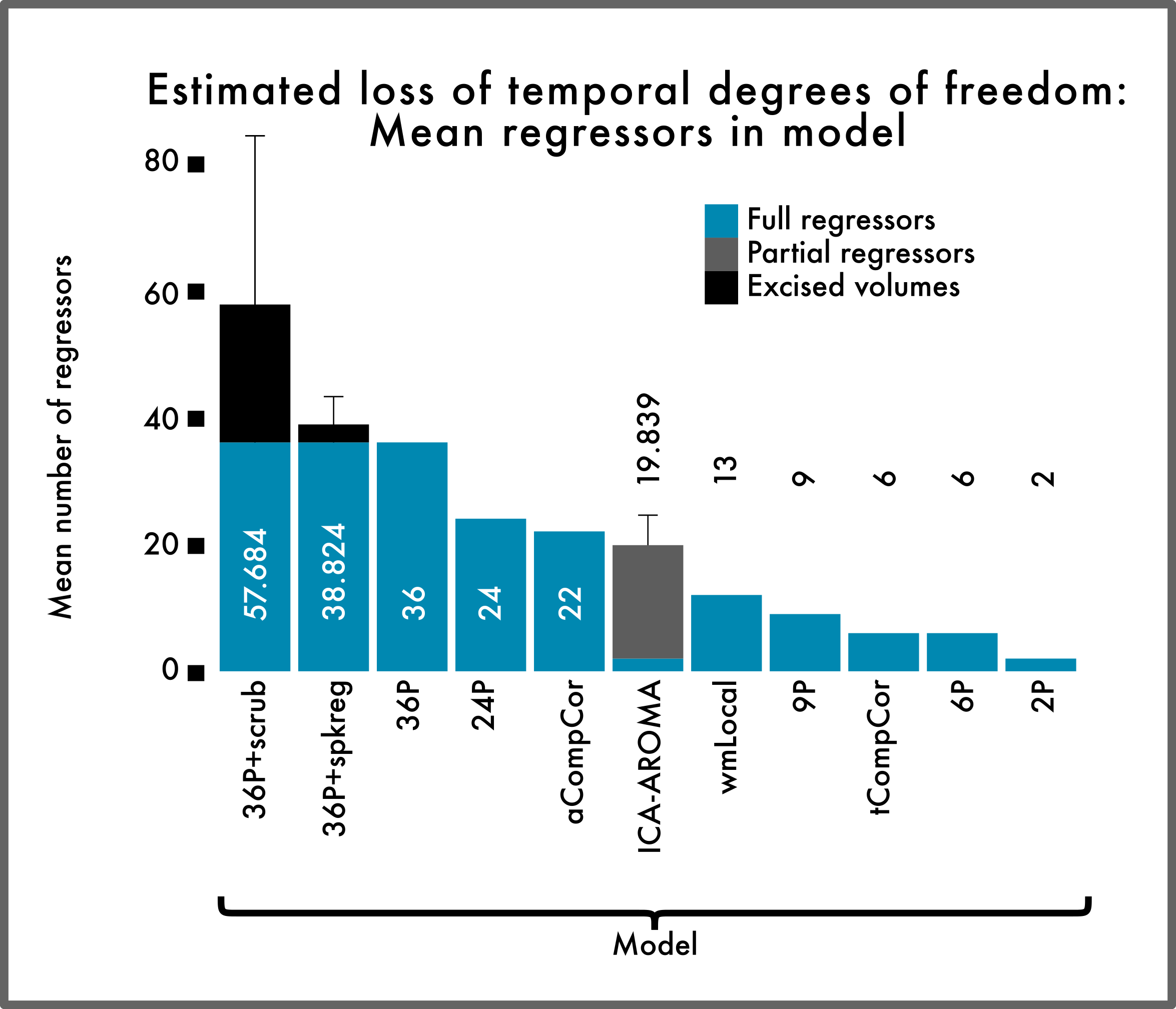}
    \caption{\textbf{Estimated loss of temporal degrees of freedom for each pipeline evaluated.}  Bars indicate mean number of additional regressors per confound model; error bars indicate standard deviation for models where the number of confound regressors varies by subject. High-parameter models and framewise censoring performed well overall on other benchmarks, but were also costliest in terms of temporal degrees of freedom.}
    \label{fig:dof}
\end{figure*}

\end{document}